\documentclass{PoS}

\usepackage{graphics,graphicx}
\usepackage{fontenc}
\usepackage{amssymb}
\usepackage{amsmath}
\usepackage{graphicx}
\usepackage{xspace}
\usepackage{natbib}
\usepackage{paralist}
\usepackage{wrapfig}
\usepackage{subcaption}
\usepackage{multicol}
\usepackage{setspace}
\usepackage{pstricks}
\usepackage{nicefrac} 
\usepackage{deluxetable}
\usepackage{caption}
\usepackage{bm}

\newcommand{\suz}{{\it Suzaku}\xspace}
\newcommand{\cen}{Cen\,X-3\xspace}
\newcommand{\xtej}{XTE\,J1946$+$274\xspace}
\newcommand{\brems}{bremsstrahlung\xspace}
\newcommand{\ms}{$M_{\odot}$\xspace}

\title{Looking into the Theory of Pulsar Accretion: \mbox{Cen X-3} and XTE J1946+274}

\ShortTitle{Looking into the Theory of Pulsar Accretion: Cen X-3 and XTE J1946+274}

\author{\speaker{Diana M. Marcu}%
 \thanks{On behalf of the Magnet Collaboration}\\
 Department of Physics \& Center for Space Science and Technology, UMBC, Baltimore, MD 21250, USA\\
 CRESST \& NASA Goddard Space Flight Center, Greenbelt, MD 20771, USA\\
 E-mail: \email{diana.m.marcu@nasa.gov}}

\author{Katja Pottschmidt\\
 Department of Physics \& Center for Space Science and Technology, UMBC, Baltimore, MD 21250, USA\\
 CRESST \& NASA Goddard Space Flight Center, Greenbelt, MD 20771, USA\\
 E-mail: \email{katja@milkyway.gsfc.nasa.gov}}

\author{Amy Gottlieb\\
 Department of Physics \& Center for Space Science and Technology, UMBC, Baltimore, MD 21250, USA\\
 CRESST \& NASA Goddard Space Flight Center, Greenbelt, MD 20771, USA\\
 E-mail: \email{amyg1@umbc.edu}}

\author{Michael T. Wolff\\
 Center for Space Research, Naval Research Laboratory, Washington DC , 20375 USA\\
 E-mail: \email{Michael.Wolff@nrl.navy.mil}}

\author{Peter A. Becker\\
 Center for Earth Observing and Space Research, George Mason University, Fairfax, VA 22030-4444, USA\\
 E-mail: \email{pbecker@gmu.edu}}

\author{J\"orn Wilms\\
 Remeis-Observatory \& ECAP, Universit\"at Erlangen-N\"urnberg, 96049 Bamberg, Germany\\
 E-mail: \email{joern.wilms@sternwarte.uni-erlangen.de}}

\author{Carlo Ferrigno\\
 ISDC, Department of Astronomy, Universit\'e de Gen\`eve, chemin d'\'Ecogia, 16, CH-1290 Versoix, Switzerland\\
 E-mail: \email{Carlo.Ferrigno@unige.ch}}

\author{Kent S. Wood\\
 Center for Space Research, Naval Research Laboratory, Washington DC , 20375 USA\\
 E-mail: \email{Kent.Wood@nrl.navy.mil}}

\abstract{This is an overview of pulsar accretion modeling. The
  physics of pulsar accretion, i.e., the process of plasma flow onto
  the neutron star surface, can be constrained from the spectral
  properties of the X-ray source. We discuss a new implementation of
  the physical continuum model developed by Becker and Wolff (2007,
  ApJ 654, 435). The model incorporates Comptonized blackbody,
  bremsstrahlung, and cyclotron emission. We discuss preliminary
  results of applying the new tool to the test cases of \suz data of
  \cen and \xtej. \cen is a persistent accreting pulsar with an O-star
  companion observed during a bright period. \xtej is a transient
  accreting pulsar with a Be companion observed during a dim
  period. Both sources show spectra that are well described with an
  empirical Fermi Dirac cutoff power law model. We extend the spectral
  analysis by making the first steps towards a physical description of
  \cen and \xtej. }

\FullConference{10th INTEGRAL Workshop: A Synergistic View of the High-Energy Sky\\
  15-19 September 2014\\
  Annapolis, MD, USA}

\begin{document}

\section{Introduction}

\begin{figure}[!t]
\begin{subfigure}{.50\textwidth}
  \centering
  \includegraphics[width=.9\linewidth]{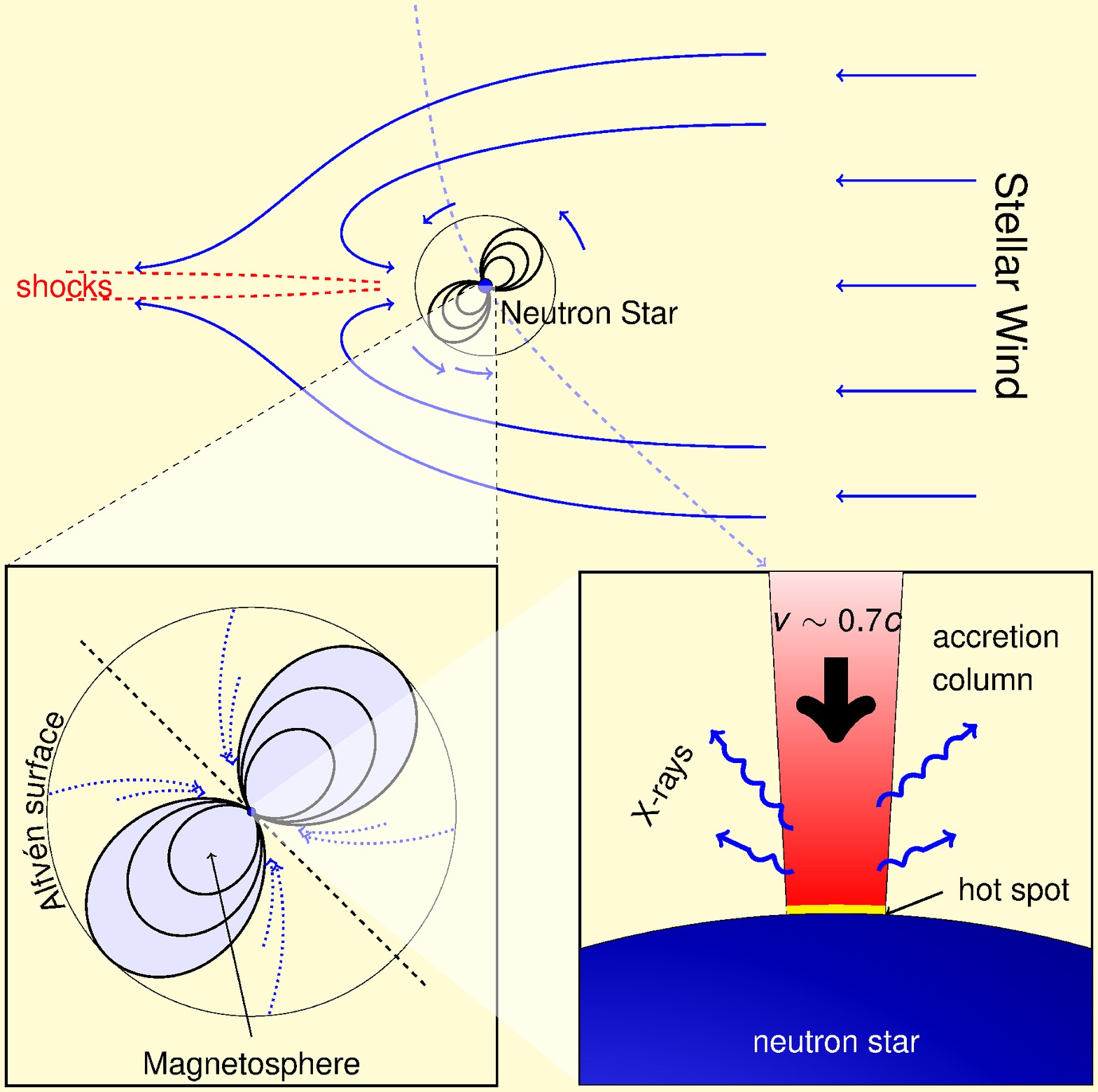}
\end{subfigure}%
\begin{subfigure}{.50\textwidth}
  \centering
  \includegraphics[width=.9\linewidth]{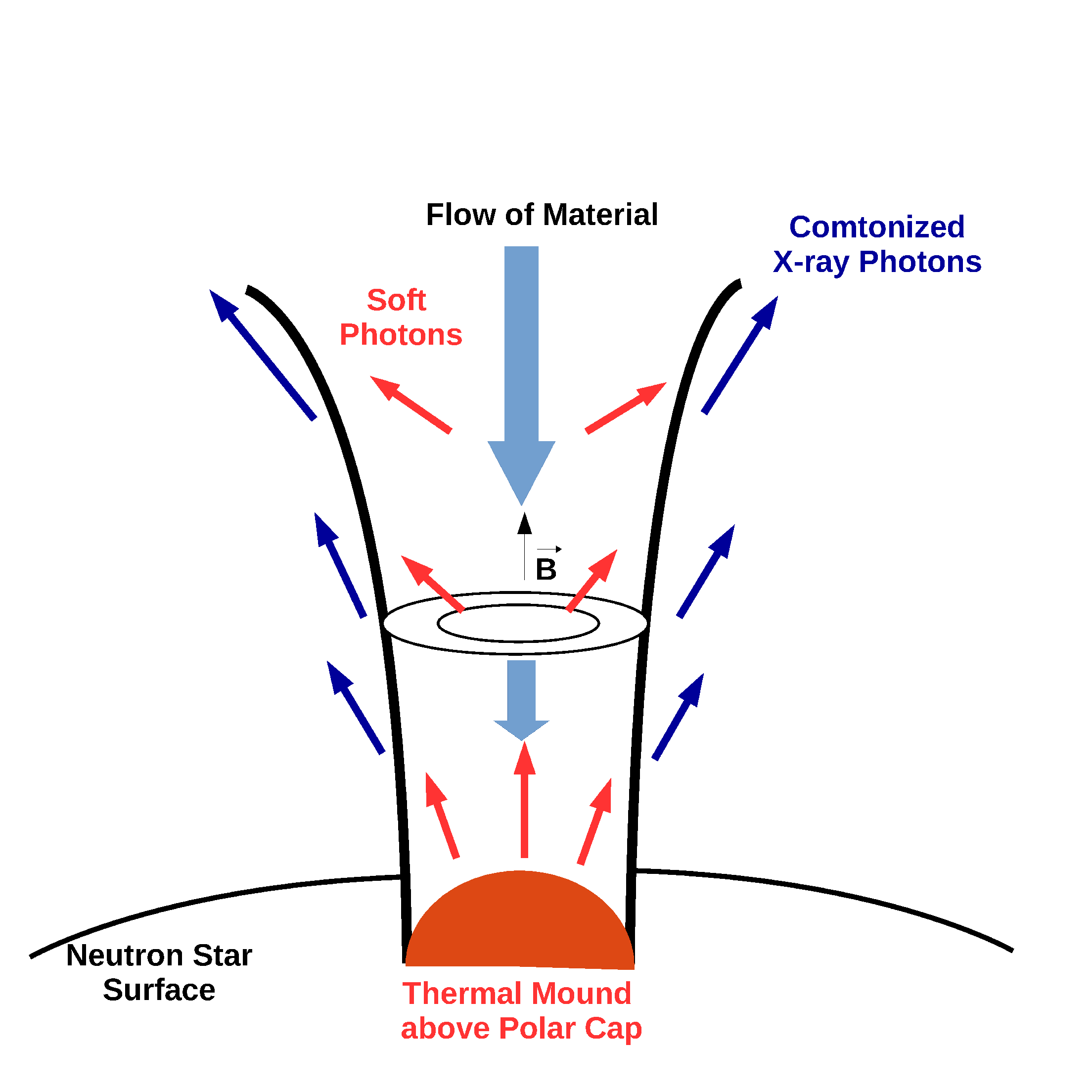}
\end{subfigure}
\caption{{\bf Left}: Schematic representation of accretion from a
  wind-emitting star onto a magnetized neutron star. After
  \cite{davidson:73}. {\bf Right}: Illustration of plasma flowing
  through the accretion column onto the magnetic pole of a neutron
  star. The soft photons are created through blackbody, \brems and
  cyclotron emission. After \cite{becker:07}.}
\label{mag_acc}
\end{figure}

X-ray pulsars are binary systems in which material from the donor star
is accreted onto a neutron star with a strong magnetic field. The
first physical continuum model for accreting pulsars was provided by
two of the coauthors, i.e., the Becker and Wolff 2007 (B\&W) model
\cite{becker:07}. The left panel in Figure\,\ref{mag_acc} illustrates
how as the material approaches the star, it couples with the magnetic
field lines at the Alfv\'{e}n radius. As the material falls towards
the surface of the neutron star, it deccelerates and produces \brems
emission. Cyclotron emission is also present due to the pulsar's
strong magnetic field. Finally, as the material settles on the surface
of the neutron star, it forms a thermal mound which emits blackbody
radiation. The B\&W model is a bulk and thermal Comptonization model
which incorporates the three types of seed photons (blackbody, \brems,
and cyclotron) that are then Comptonized into the approximate power
law spectrum that we see in accreting pulsars, see right panel in
Figure\,\ref{mag_acc}. We are working on implementing this model for
statistical fitting, e.g. in {\tt xspec} or {\tt isis}, and making it
publicly available. Currently phenomenological models are generally
used to describe the cutoff power law continuum shape of accreting
pulsars.

\section{Observations}

We will first introduce the sources, \cen and \xtej, and discuss
empirical modeling of their \suz spectra. We will then present the
fist results of applying the B\&W model to these spectra.

\begin{figure}[!t]
\begin{subfigure}{.50\textwidth}
  \centering
  \includegraphics[width=.9\linewidth]{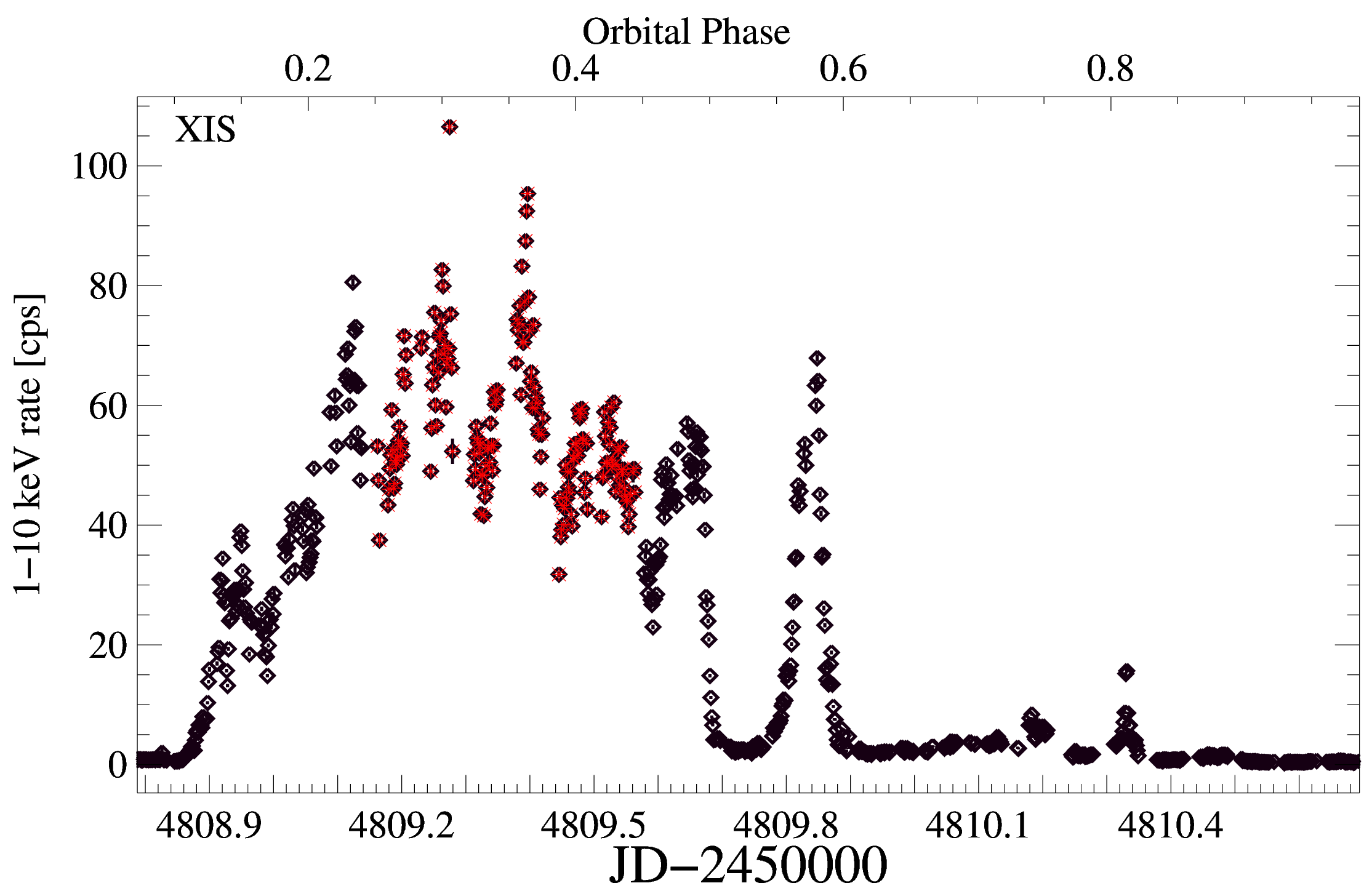}
\end{subfigure}%
\begin{subfigure}{.50\textwidth}
  \centering
  \includegraphics[width=.9\linewidth]{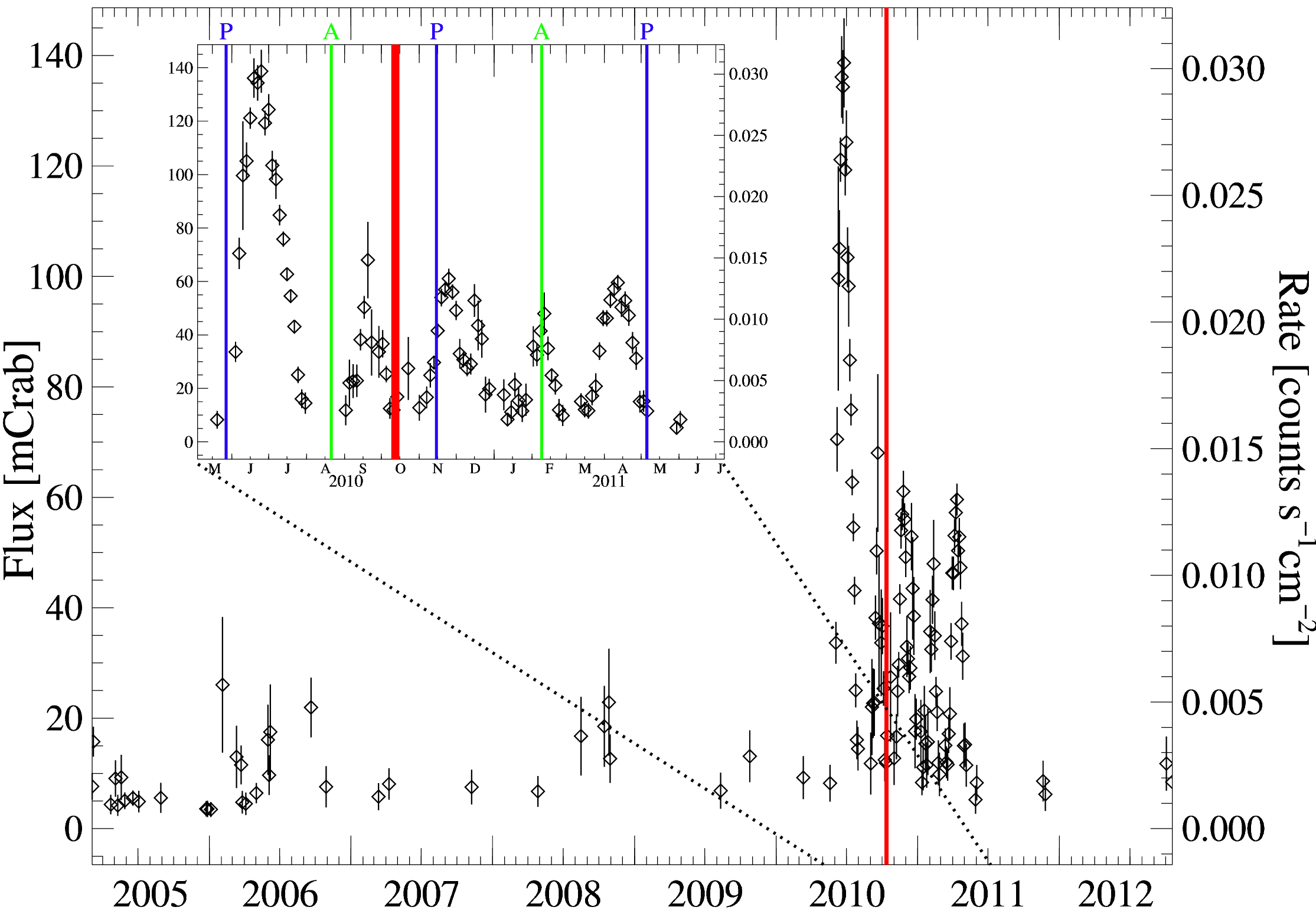}
\end{subfigure}
\caption{{\bf Left} - \suz-XIS lightcurve of an observation of a full
  \cen orbit in 2008, December. The spectral analysis was performed on
  the data highlighted in red (after Gottlieb et al., in prep.). {\bf
    Right} - \textsl{Swift}-BAT lightcurve of \xtej with the time of
  the \suz observation marked as the red vertical line. The upper-left
  panel shows a closer view of the 2010 series of outbursts with
  apastron and periastron times (after Marcu et al., in prep.).}
\label{lc}
\end{figure}

\subsection{Centaurus X-3}

\cen was the first X-ray pulsar discovered in a 1971 observation, with
     {\sl Uhuru} \cite{giacconi:71}. \cen is a persistent High Mass
     X-ray Binary (HMXB) composed of a neutron star with a mass of
     $1.21\pm0.21$\,\ms and a pulse period of $\sim4.8$\,s which
     orbits an O6.5II-type companion with a mass of $20.5\pm0.7$\ms
     \cite{ash:99}. The X-ray source is eclipsed $\sim20\%$ of its
     2.1\,d orbit. \cen also exhibits several spectral features: three
     iron (Fe) fluorescent lines at 6.4\,keV, 6.7\,keV and 6.9\,keV
     \cite{naik:11}, a Cyclotron Scattering Resonance Feature (CRSF)
     at $\sim30$\,keV \cite{suchy:08}, and a broad ``10\,keV bump'' at
     $\sim13$\,keV, a feature which is relatively common in accreting
     pulsars \cite{coburn:02,mihara:95}.
 
\begin{figure}[!t]
\begin{subfigure}{.50\textwidth}
  \centering
  \includegraphics[width=.9\linewidth]{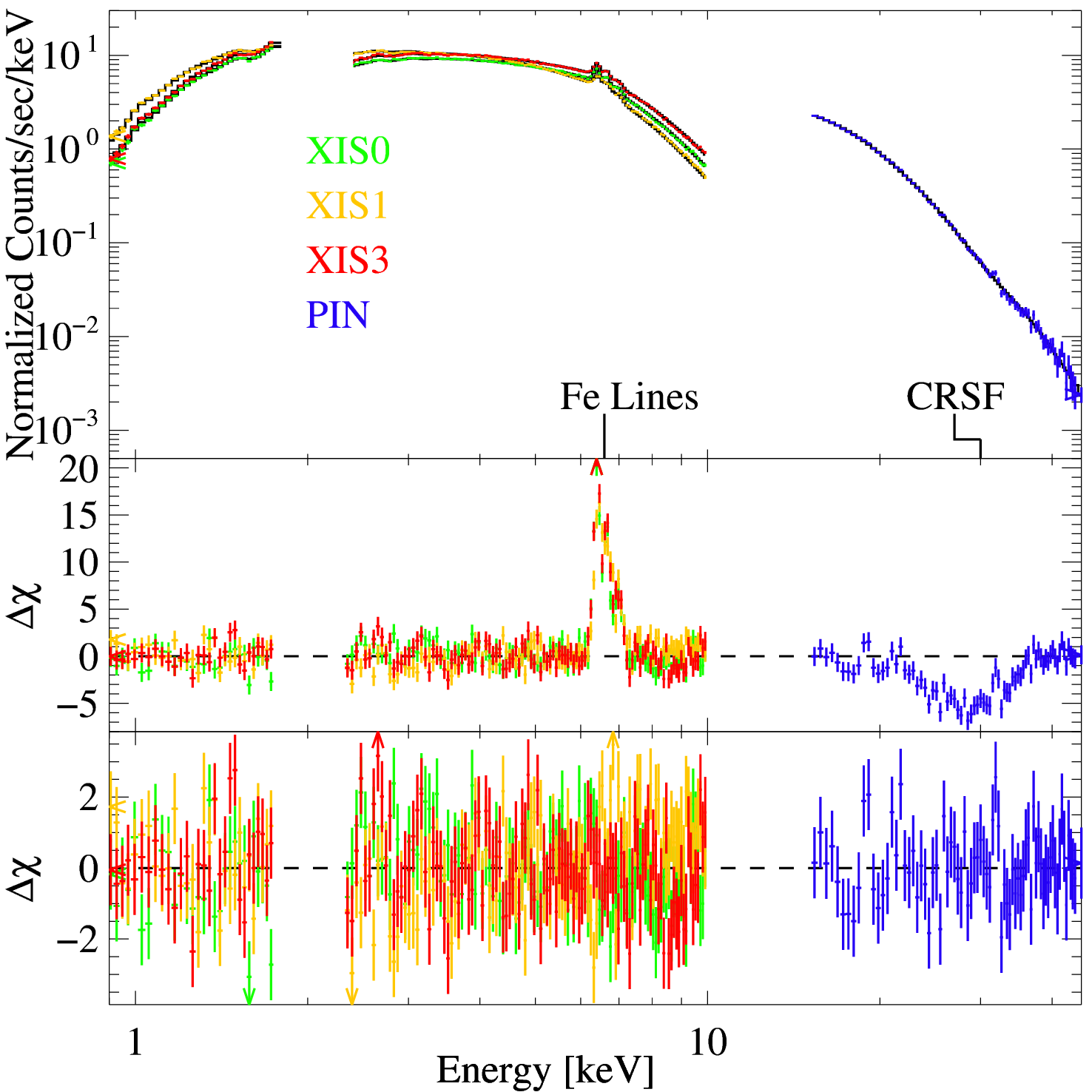}
\end{subfigure}%
\begin{subfigure}{.50\textwidth}
  \centering
  \includegraphics[width=.9\linewidth]{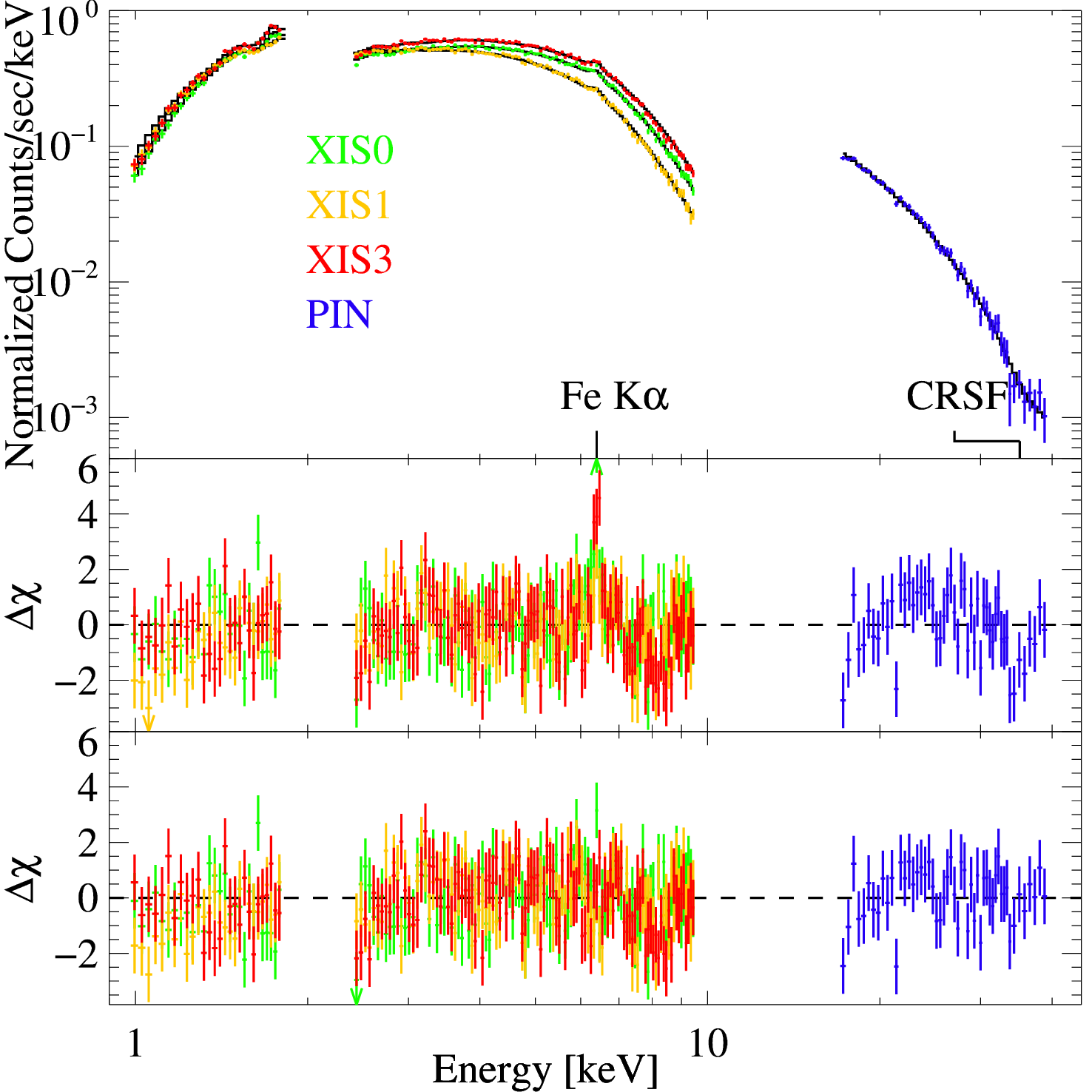}
\end{subfigure}
\caption{Top: \suz spectra of the \cen data highlighted in red in the
  left panel of Figure\,2 ({\bf left}) and \suz spectra of \xtej ({\bf
    right}) with the Fermi Dirac cutoff power law best fit model
  (black line). Middle: Residuals of the fitted absorbed continuum
  without the Fe line and CRSF. Bottom: Residuals after including the
  Fe Gaussian line and CRSF.  Selected best fit parameters are listed
  in Table\,1.}
\label{obs_spec}
\end{figure}

We studied a 90\,ks \suz observation spread out over a full orbital
period observed in 2008, December $8-10$, eclipse to eclipse, as can
be seen in the left panel of Figure\,\ref{lc}. The spectra shown in
the left panel of Figure\,\ref{obs_spec} were extracted from a bright
part of the lightcurve (shown in red in the left panel of
Figure\,\ref{lc}) in which the hardness ratio was approximately
constant. Considering a distance to the source of $\sim8$\,kpc
\cite{krzeminski:74}, the $1-40$\,keV luminosity for the extracted
bright part is $\sim1.2\times 10^{38}$\,erg/s.

We modeled the XIS and PIN spectra (Gottlieb et al., in prep.) by
taking into account the following components in {\tt xspec}: the
cross-normalization between the instruments modeled with a {\tt
  constant}, the partial covering absorption modeled with {\tt
  tbnew\_pcf} and {\tt tbnew\_feo}\footnote{Tbnew is an updated
  version of tbabs, see
  http://pulsar.sternwarte.uni-erlangen.de/wilms/research/tbabs/ for
  more details.} \cite{wilms:00}, the three Fe emission lines modeled
with Gaussian lines ({\tt gauss}), the 13\,keV feature modeled with
another Gaussian line, and a CRSF at $\sim30$\,keV modeled with a
Gaussian optical depth profile ({\tt gabs})\footnote{Three additional
  Gaussian lines were included to model weak residuals. They will be
  discussed in Gottlieb et al., in prep.}. The continuum was best
fitted with a Fermi Dirac cutoff model \cite{tanaka:86} described by:
\begin{equation}
M_\mathrm{fdcut}(E)\propto E^{-\Gamma}\times\left[1+\exp\left(\frac{E-E_\mathrm{cut}}{E_\mathrm{fold}}\right)\right]^{-1}.
\end{equation}
At energy $E$ the photon flux is described by a power law with a
photon index $\Gamma$, multiplied by an exponential cutoff at energy
$E_{\mathrm{cut}}$ with a folding energy $E_{\mathrm{fold}}$. The
final best fit model is:
\begin{eqnarray}\nonumber
M_\text{Cen\,X-3}(E) = \texttt{const} \times \texttt{tbnew\_pcf} \times \texttt{tbnew\_feo} \times (\texttt{power} \times \texttt{fdcut} \\
+ 3 \times \texttt{gauss}_\mathrm{Fe} + \texttt{gauss}_\mathrm{10\,keV} + 3 \times \texttt{gauss}_\mathrm{weak}) \times \texttt{gabs}_\mathrm{CRSF}.
\label{eq:cenx3_model}
\end{eqnarray}
Selected parameters of this empirical fit are listed in
Table\,\ref{emp}. The continuum parameters are highlighted. Those are
the standard model parameters that describe the continuum which we are
trying to describe with the physical B\&W model.

Applying an analytical test implementation of the full B\&W model (in
Fortran), we performed a fit ``by eye'' to the {\bf unfolded,
  unabsorbed} \suz data of \cen. The result is shown in the left panel
of Figure\,\ref{sim_spec} with the physical parameters listed in
Table\,\ref{phys}. The results are promising since the model
qualitatively agrees with the data. The emission is \brems dominated
overall, with a strong contribution from cyclotron emission. The CRSF
is not implemented in the model and, therefore, we can see the model
overpredicting the flux around 30\,keV where the observed spectrum
shows a broad dip due to the magnetic resonant scattering of the
photons. The Fe fluorescent lines, which are also not part of the
continuum model, can be seen in the 6-7\,keV region.

\subsection{\xtej}

\xtej was discovered in 1998 September by the All-Sky Monitor (ASM),
on board of the Rossi X-ray Timing Explorer ({\it RXTE)}
\cite{smith:98}. In contrast to the bright and persistent \cen, \xtej
is a transient HMXB which has shown only two series of outbursts:
three months in 1998 at discovery and several months in 2010-2011. The
optical companion is a giant Be-IV-IVe star \cite{verrecchia:02}
orbited by a $\sim15.7$\,s pulsar \cite{muller:12}. The orbital period
is $\sim169.2$\,d, the orbital inclination is $\sim46^{\circ}$ and the
distance to the source is $9.5\pm2.9$\,kpc \cite{wilson:03}. The
source shows $\sim2$ outbursts per orbit, an unusual behavior for Be
systems in which only one outburst per orbit is generally
observed. During the 1998 series a CRSF was detected at $\sim35$\,keV
\cite{heindl:01}. \xtej also shows an Fe fluorescent line at 6.4\,keV.

We extracted \suz XIS and PIN spectra from a 42\,ks observation taken
on 2010 October 12, during a minimum between two outburst from the
2010 series (see right panel of Figure\,\ref{lc}). During the
observation the source had a $1-40$\,keV luminosity of $\sim3.4\times
10^{36}$\,erg/s (much lower than that of \cen). Similarly to \cen, we
modeled the XIS and PIN data (Marcu et al., in prep.)  taking into
consideration the instrument cross calibration, absorption, a Gaussian
Fe line, and a weak CRSF line. Even though these sources have very
different characteristics (e.g. luminosity, optical companion,
frequency of bright periods, orbital periods, etc.) their continua are
well described with the same empirical model, {\tt
  fdcut}\footnote{Note that this empirical description is not unique,
  see Marcu et al., in prep. for the application of all commonly used
  empirical models, yielding comparable fit quality.}. The best fit
model for \xtej in {\tt xspec} notation is:
\begin{equation}
M_\text{XTE\,J1946+274}(E) = \texttt{const} \times \texttt{tbnew\_feo} \times (\texttt{power} \times \texttt{fdcut} + \texttt{gauss}_\mathrm{Fe}) \times \texttt{gabs}_\mathrm{CRSF}.
\label{eq:xtej_model}
\end{equation}
Selected parameters for this fit are listed in Table\,\ref{emp}. The
continuum paramteters for \xtej are also highlighted.

\begin{figure}[!t]
\begin{subfigure}{.50\textwidth}
  \centering
  \includegraphics[width=.9\linewidth]{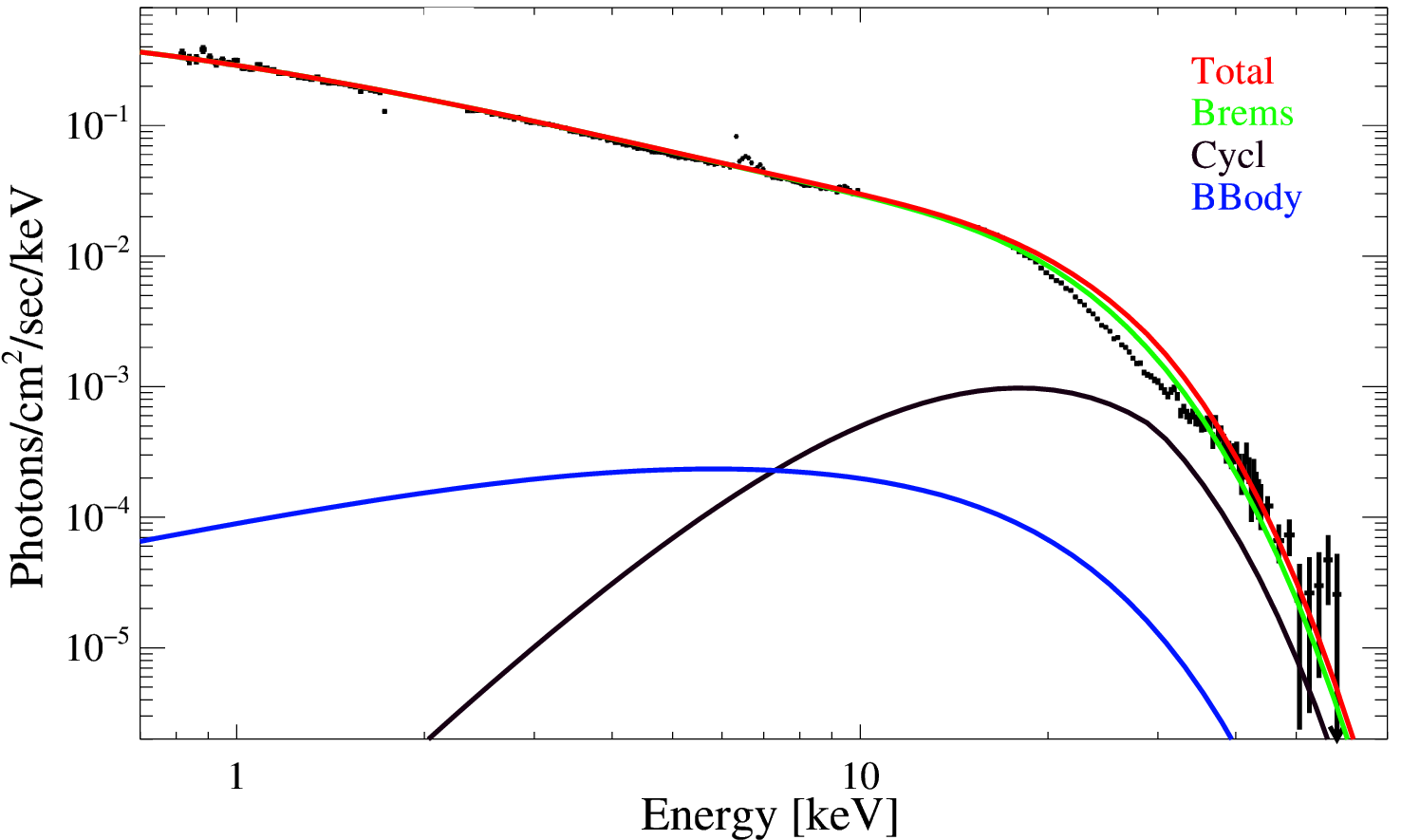}
\end{subfigure}%
\begin{subfigure}{.50\textwidth}
  \centering
  \includegraphics[width=.9\linewidth]{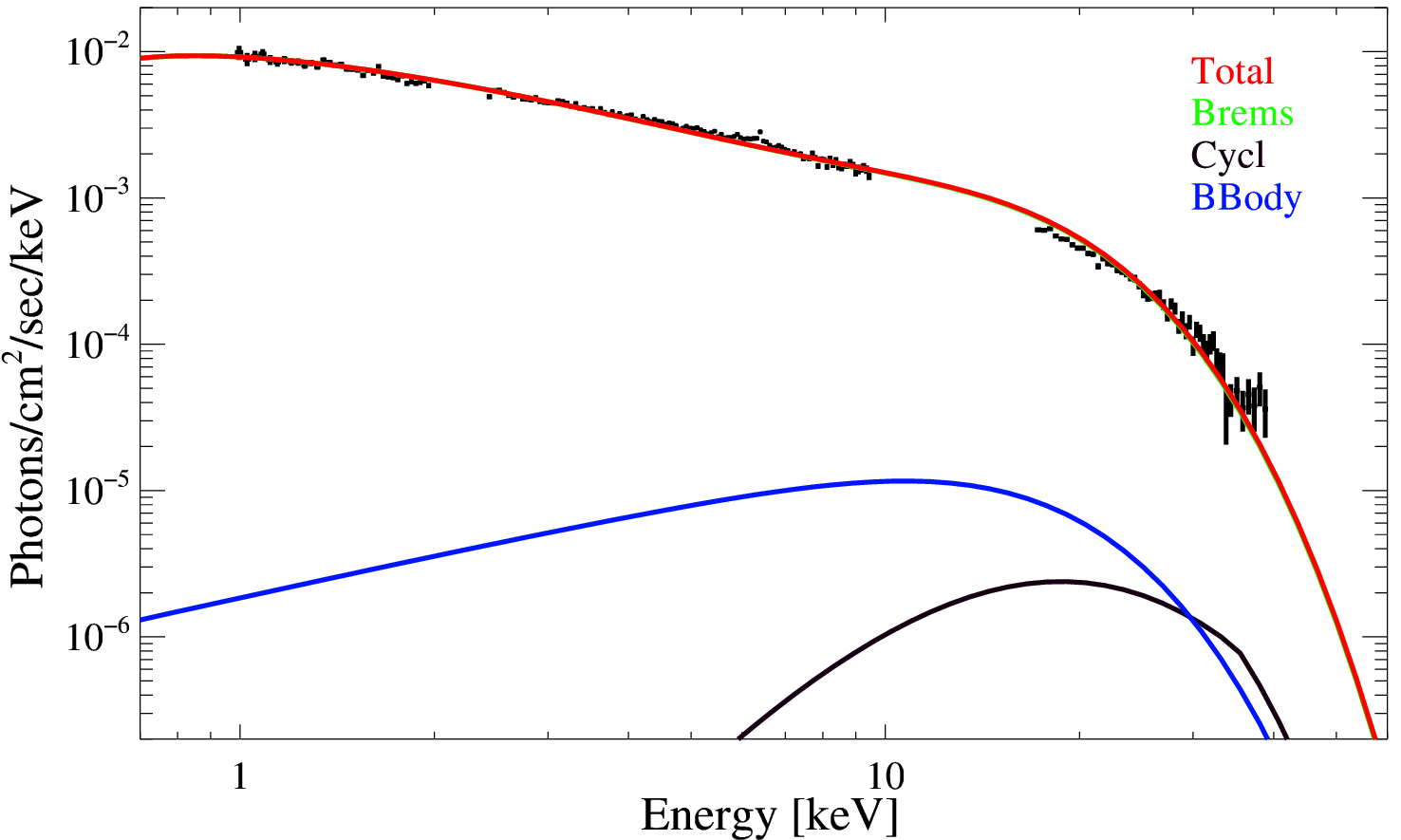}
\end{subfigure}
\caption{Fit ``by eye'' applying the B\&W physical model to the \suz
  data (black) of \cen ({\bf left}) and \xtej ({\bf left}). The three
  types of Comptonized emission are shown: \brems in green, cyclotron
  emission in dark purple and back body emission in blue. The physical
  parameters are listed in Table\,3.}
\label{sim_spec}
\end{figure}

Similarly to \cen, we performed a ``fit by eye'' to the unfolded,
unabsorbed \suz data of \xtej using the new B\&W model
implementation. As can be seen in the right panel in
Figure\,\ref{sim_spec}, the continuum is again dominated by \brems
emission, but with a much weaker contribution from cyclotron line
emission than we found in \cen. The model again agrees well with the
data, particularly at lower energies. The physical model parameters
applied towards this ``fit'' are listed in Table\,\ref{phys}. As
expected, due to the difference in luminosity between the two sources,
the mass accretion rate, electron temperature inside the column, as
well as the radius of the accretion column are lower for \xtej than
they are for the high-luminosity source \cen.

\section{Current and Future Applications}

An earlier application of the B\&W model was carried out for the
source 4U\,0115$+$63 using broad-band \textsl{BeppoSAX} data
accumulated during an outburst decay, see \cite{ferrigno:09} and
\texttt{xspec} model available from C.\ Ferrigno. In this case the
emission from the column is dominated by Compton scattered cyclotron
emission because of the relatively low magnetic field of this
source. Its contribution is dominant above $\sim$7\,keV. Below this
energy, it was necessary to include a thermal Comptonization
component, which is interpreted as coming from a scattering halo
surrounding the neutron star surface.

The ``\texttt{compMag}'' model \cite{farinelli:12} provides a
numerical solution of the radiative transfer equation in a cylindrical
column with approximate cross sections for scattering in a magnetized
plasma using similar assumptions as the B\&W model. In the version
available for Xspec, only blackbody seed photons are taken into
account in an exponential distribution along the accretion
column. This model allows the user to switch between two different
velocity profiles of the in falling plasma. Work is in progress to
include emission from the magnetized optically thin plasma of the
accretion stream and will be presented in a forthcoming publication,
see also \cite{ferrigno:14}.

Here we presented promising first steps towards the physical modeling
of accreting pulsars using a new analytical implementation of the B\&W
model. In order to determine the robustness of the physical parameters
statistical modeling is required and we are currently implementing the
analytical model into \texttt{xspec}. The analytical and numerical
versions of the model will be tested on observed X-ray spectra of
several accreting pulsars starting with \cen, \xtej, 4U\,0115$+$63 and
GX\,304$-$1. We plan to conduct a thorough comparison between the
empirical and physical fits. This will allow us to physically
interpret data of accreting pulsars previously analyzed with empirical
models.

\begin{table}
\parbox{.48\linewidth}{
\centering
\begin{tabular}{lcc}
  \hline
  \hline
  &\cen & \xtej \\
  \hline\noalign{\smallskip}
  $N_\mathrm{H}^a$ &  $4.8^{+2.3}_{-1.3}$ & - \\
  $f$ & $0.26^{+0.10}_{-0.06}$  & - \\
  $N_\mathrm{H}^a$ & $1.61^{+0.07}_{-0.05}$ & $1.67(3)$\\
  $A_\Gamma^b$ & $0.37^{+0.09}_{-0.03}$ & $2.05^{+0.04}_{-0.05}$ $^c$\\
  $\bm{\Gamma}$ & $\bm{1.06^{+0.16}_{-0.05}}$ & $\bm{0.57(2)}$ \\
  $\bm{E_\mathrm{cut}}$\,[keV] & $\bm{17.7^{+2.7}_{-3.3}}$ &  $\bm{0.05^{+0.03}_{-0.05}}$ $^c$ \\
  $\bm{E_\mathrm{fold}}$\,[keV] & $\bm{7.0^{+0.07}_{-0.06}}$ &  $\bm{8.9(4)}$  \\
  $E_{\mathrm{CRSF}}$\,[keV] & $30.3(9)$ & $35.2^{+1.5}_{-1.3}$ \\
  ${\chi}^{2}_{\mathrm{red}}/\mathrm{dof}$ & $1.31/500$ & $1.17/466$\\
  \hline   
\end{tabular}
\caption{Main best fit model parameters for the Fermi Dirac cutoff
  power law fits to the \suz spectra of \cen and \xtej. The continuum
  parameters are highlighted. The full list of fit parameters can be
  found in Gottlieb et al., in prep. for \cen and Marcu et al., in
  prep. for \xtej. (a)$\times10^{22}\mathrm{cm}^{-2}$;
  (b) keV$^{-1}$\,cm$^{-2}$\,s$^{-1}$; (c)$\bm{\times 10^{-2}}$.}
\label{emp}}
\hfill
\parbox{.48\linewidth}{
\centering
\begin{tabular}{lcc}
  \hline
  \hline
  &\cen & \xtej \\
  \hline\noalign{\smallskip}
  $\dot{M}\,[\times 10^{17}\,\mathrm{g/s}]$ & 11.32 & 1.4 \\ 
  $T_\mathrm{e}\,[\times 10^7\,\mathrm{K}]$ & 3.74 & 3.4 \\
  $B\,[\times 10^{12}\,\mathrm{G}]$ & 2.55 & 3.04 \\
  $r_0\,[\mathrm{m}]$ & 700 & 100 \\
  $\sigma_\parallel^d$ & 1.01 & 2.141 \\
  $\bar{\sigma}^d$ & 9.317 & 12.79 \\
\hline
\end{tabular}
\caption{Physical parameters used for a ``fit by eye'' of the B\&W
  model for the \suz data of \cen and \xtej. The physical parameters
  are: mass accretion rate, $\dot{M}$; electron temperature inside the
  column, $T_\mathrm{e}$; magnetic field strength in the emission line
  region, $B$; radius of the accretion column, $r_0$; the scattering
  cross sections parallel to the magnetic field ($\sigma_\parallel$)
  and the average cross section ($\bar{\sigma}$) as a function of the
  Thompson cross section, $\sigma_\mathrm{T}$.  (d)$\times
  10^{-4}\,\sigma_\mathrm{T}$.}
\label{phys}}
\end{table}

\newpage

%\bibliographystyle{jwapjbib}
%\bibliography{mnemonic,ms}

\end{document}